\documentclass[english,prb,preprint,longbibliography]{revtex4-1}
\usepackage[T1]{fontenc}
\usepackage[latin9]{inputenc}
\usepackage{amsmath}
\usepackage{graphicx}

\makeatletter

\providecommand{\tabularnewline}{\\}

\usepackage{graphicx}
\usepackage{slashed}
\usepackage{bbm}
\usepackage{tikz}
\usepackage{tikz-3dplot}

\makeatother

\usepackage{babel}
\begin{document}
\title{Massless Majorana bispinors and two-qubit entangled states}
\author{R. Romero}
\email{rromero@correo.cua.uam.mx}

\address{Departamento de Ciencias Naturales, Universidad Aut\'onoma Metropolitana-Cuajimalpa,
05348 Ciudad de M\'exico, M\'exico.}
\begin{abstract}
This is a pedagogical paper, where bispinors solutions to the four-dimensional
massless Dirac equation are considered in relativistic quantum mechanics
and in quantum computation, taking advantage of the common mathematical
description of four-dimensional spaces. First, Weyl and massless Majorana
bispinors are shown to be unitary equivalent, closing a gap in the
literature regarding their equivalence. A discrepancy in the number
of linearly independent solutions reported in the literature is also
addressed. Then, it is shown that Weyl bispinors are algebraically
equivalent to two-qubit direct product states, and that the massless
Majorana bispinors are algebraically equivalent to maximally entangled
sates (Bell states), with the transformations relating the two bispinors
types acting as entangling gates in quantum computation. Different
types of entangling gates are presented, highlighting a set that fulfills
the required properties for Majorana zero mode operators in topological
quantum computation. Based on this set, a general topological quantum
computation model with four Majorana operators is presented, which
exhibits all the required technical and physical properties to obtain
entanglement of two logical qubits from topological operations.
\end{abstract}
\maketitle

\section{Introduction}

A Majorana fermion is a spin 1/2 particle that is its own antiparticle.
They were first proposed in 1937 by E. Majorana\citep{Majorana2008}
in the context of particle physics. As an elementary particle, the
only fundamental candidate for a Majorana fermion is the massive neutrino.
It could also be a Dirac particle, although the Majorana alternative
is theoretically preferred.\citep{CBO9781107358362A095,Mohapatra:2005wg}
The experimental verification of the Majorana nature of the neutrino,
through the observation of neutrinoless double beta decay processes,
is still an open question. 

Majorana fermions arise also in condensed matter systems.\citep{0034-4885-75-7-076501,0268-1242-27-12-124003,doi:10.1146/annurev-conmatphys-030212-184337,Sarma2015}
Here they are not elementary particles, but rather localized zero-energy
bound states (Bogoliubov quasiparticles) of electrons and holes, better
known as Majorana zero modes\footnote{As is nicely put in Ref. \onlinecite{0268-1242-27-12-124003}, an
MZM is \textquotedbl\emph{a quasiparticle that is its own hole}.\textquotedbl} (MZMs). In this case the Majorana condition is satisfied through
the use of Hermitian operators to describe MZMs. The composite objects
consisting of Majorana bound states coupled to topological defects,
such as vortices, obey non-Abelian statistics and are known as Ising
anyons,\citep{MOORE1991362,PhysRevA.73.042313} which constitute a
particular type of non-Abelian anyons. Examples of 2-d systems admitting
Ising anyons are the $\nu=5/2$ fractional quantum Hall state,\citep{MOORE1991362,NAYAK1996529}
$p+ip$ superconductors,\citep{PhysRevLett.86.268,PhysRevB.61.10267}
and the surface of topological insulators,\citep{PhysRevLett.100.096407}
among others.

The interest in Ising anyons, from the perspective of quantum computation,
is because they provide a means for fault-tolerant quantum computation.\citep{Kitaev20032,key1943131m,RevModPhys.80.1083,Sarma2015}
In a system with localized anyons quantum information can be stored
non-locally in pairs, or in general $n$-tuplets, with $n$ even,
of anyons. Computations are performed by adiabatically braiding the
anyons worldlines. These braiding operations constitute the logical
quantum gates acting on the states and, up to a phase, depend only
on the topology of the trajectories, in turn classified by the braid
group. A topological quantum computation (TQC) model is specified\citep{PhysRevA.73.042313}
by providing the Hilbert space, the initial state, the braid operators
and the measurable observables. 

It has been shown that the operators representing the MZMs can be
given in terms of Dirac gamma matrices\citep{NAYAK1996529,JACKIW1981681,PhysRevB.85.033102}
and, in particular, in Refs. \onlinecite{JACKIW1981681} and \onlinecite{PhysRevB.85.033102}
it is shown that the Clifford algebra of the Majorana operators, for
a 2-d system with four vortices, can be realized by elements of the
4-d spacetime Clifford algebra. This result suggests that a common
mathematical description can be given for the four-component spinors
(bispinors) and the relevant particle states in TQC, namely Weyl\footnote{Weyl spinors are always massless, and in the condensed matter literature
massless fermions are usually referred to as Dirac fermions, but being
massless they are indeed Weyl fermions.} and massless Majorana states.

In this paper we study massless Majorana bispinors, that is solutions
to the 4-d massless Dirac equation satisfying the Majorana condition,
in two different settings: relativistic quantum mechanics (RQM) and
quantum computation (QC). In the first instance, besides showing explicit
general solutions to the equation, which are difficult to find in
the literature, if at all, we complete the known equivalence between
massless Majorana and Weyl free field operators by showing that it
also holds for $c$-number bispinors. 

We also address an inconsistency in the number of linearly independent
solutions to the massless Dirac equation reported in the literature
where, up to differences in normalization and sign factors, it is
stated that positive and negative energy solutions are proportional
in momentum space.\citep{itzykson2006quantum,schwabl2008advanced,srednicki2007quantum,Dixon:1996wi}
This statement is made in terms of four-component Dirac spinors, and
it then follows that only two linearly independent bispinor solutions
exist in the massless case. Taking the limit $m\rightarrow0$ in massive
solutions, as is done in Refs. \citep{Lurie1968,gross1999relativistic,das2008lectures},
also leads to an incomplete set of solutions. However, from a pure
mathematical viewpoint, the massless Dirac equation in momentum space
is an algebraic equation, whose solutions are the eigenvectors of
a $4\times4$ Hermitian matrix, so that four independent solutions
must exist. Indeed, this result is also found in the literature for
the special case of momentum along the direction of $\hat{\mathbf{z}}$.\citep{schweber2005introduction,greiner2000relativistic}

In the QC context, we establish an algebraic equivalence between Weyl
bispinors and bipartite qubit states. We show that the unitary transformations
relating the Weyl and Majorana bispinors in RQM play the role of entangling
two-qubit gates in QC, and that maximally entangled states (Bell states)
are algebraically equivalent to massless Majorana bispinors.\footnote{In Ref. \onlinecite{PhysRevA.64.024303} it is already established
a connection between entanglement and invariance under charge conjugation
for two-qubit density matrices} Different types of entangling gates are discussed, providing a list
not meant to be exhaustive. A set of the entangling gates fulfills
the requirements for MZMs operators, and we use it to construct a
TQC toy model with four MZMs from the bottom up, showing how to obtain
operators and states, as well as entanglement of two logical qubits
from braiding. 

The organization is as follows: In section II we obtain bispinor solutions
to the massless Dirac equation and show that they are unitarily equivalent.
The completeness of the solutions is also discussed. In section III
we establish the algebraic equivalence between massless bispinors
and two-qubit states and discuss the entangling gates. In section
IV we provide a TQC toy model based on a set of the entangling gates.
Finally, concluding remarks are given.

\section{Massless \emph{c}-number bispinors}

\subsection{Weyl }

Let us begin by considering four-component Weyl bispinors with four-momentum
$p^{\mu}=\left(\pm\left|\mathbf{p}\right|,\mathbf{p}\right)$, respectively
for positive- and negative-energy $p^{0}=\pm E=\pm\left|\mathbf{p}\right|$,
which are solutions to the massless Dirac equation

\begin{equation}
i\gamma^{\mu}\partial_{\mu}\Psi=0.\label{eq:2.1}
\end{equation}

\noindent The gamma matrices $\gamma^{\mu}=\left(\gamma^{0},\boldsymbol{\gamma}\right)$
obey the Clifford algebra relation

\noindent 
\begin{equation}
\gamma^{\mu}\gamma^{\nu}+\gamma^{\nu}\gamma^{\mu}=2g^{\mu\nu},\label{eq:2.1-1}
\end{equation}

\noindent with $g^{\mu\nu}$ the metric tensor with signature $\text{diag}(1,-1,-1,-1)$,
and the Weyl representation

\begin{equation}
\begin{array}{c}
\gamma^{0}=\left(\begin{array}{rr}
0 & 1\\
1 & 0
\end{array}\right),\,\boldsymbol{\gamma}=\left(\begin{array}{rr}
0 & \boldsymbol{\sigma}\\
-\boldsymbol{\sigma} & 0
\end{array}\right),\end{array}\label{eq:2.2}
\end{equation}

\noindent with $\boldsymbol{\sigma}=\left(\sigma^{1},\sigma^{2},\sigma^{3}\right)$
the standard Pauli matrices will be used throughout. Using the plane
waves

\begin{equation}
\Psi=u(\mathbf{p})\exp\left\{ i\left(\pm Et-\mathbf{x}\cdot\mathbf{p}\right)\right\} ,\label{eq:2.3}
\end{equation}

\noindent and the matrices

\begin{equation}
\begin{array}{c}
\gamma^{5}\equiv i\gamma^{0}\gamma^{1}\gamma^{2}\gamma^{3}=\left(\begin{array}{rr}
-1 & 0\\
0 & 1
\end{array}\right),\,\boldsymbol{\Sigma}\equiv\gamma^{5}\gamma^{0}\boldsymbol{\gamma}=\left(\begin{array}{rr}
\boldsymbol{\sigma} & 0\\
0 & \boldsymbol{\sigma}
\end{array}\right),\end{array}\label{eq:2.4}
\end{equation}

\noindent equation (\ref{eq:2.1}) is rewritten as

\begin{equation}
\boldsymbol{\Sigma}\cdot\mathbf{\hat{p}}\,u(\mathbf{p})=\pm\gamma^{5}\,u(\mathbf{p}),\label{eq:2.5}
\end{equation}

\noindent with $\mathbf{\hat{p}}=\mathbf{p}/\left|\mathbf{p}\right|$.
Thus, the bispinors $u(\mathbf{p})$ are eigenvectors of both helicity
$\boldsymbol{\Sigma}\cdot\mathbf{\hat{p}}$ and chirality $\gamma^{5}$
operators, and Eq. (\ref{eq:2.5}) expresses the known result that
chirality equals the helicity for massless, positive-energy bispinors,
while it is opposite for negative-energy ones. Taking the direction
of $\mathbf{p}$ along $\hat{\mathbf{z}}$ (from now on called the
canonical frame) in Eq. (\ref{eq:2.5}) one obtains the four independent
solutions,\footnote{\noindent This set of solutions is reported in Refs. \onlinecite{greiner2000relativistic,schweber2005introduction,Grensing:1563285},
with variations due to the gamma matrices representation employed.} with their eigenvalues given in Table 1.

\begin{equation}
\begin{array}{rr}
u^{(1)}\left(p_{z}\right)=\begin{pmatrix}0\\
0\\
1\\
0
\end{pmatrix}, & u^{(2)}\left(p_{z}\right)=\begin{pmatrix}0\\
1\\
0\\
0
\end{pmatrix},\\
\\
u^{(3)}\left(p_{z}\right)=\begin{pmatrix}0\\
0\\
0\\
1
\end{pmatrix}, & u^{(4)}\left(p_{z}\right)=\begin{pmatrix}1\\
0\\
0\\
0
\end{pmatrix}.
\end{array}\label{eq:2.6}
\end{equation}

\begin{table}
\begin{centering}
\begin{tabular}{|c|c|c|c|c|}
\hline 
 & $u^{(1)}\left(p_{z}\right)$ & $u^{(2)}\left(p_{z}\right)$ & $u^{(3)}\left(p_{z}\right)$ & $u^{(4)}\left(p_{z}\right)$\tabularnewline
\hline 
\hline 
Energy & + & + & - & -\tabularnewline
\hline 
Helicity & 1 & -1 & -1 & 1\tabularnewline
\hline 
Chirality & 1 & -1 & 1 & -1\tabularnewline
\hline 
\end{tabular}
\par\end{centering}
\caption{Eigenvalues of the canonical frame Weyl bispinors}
\end{table}

To obtain solutions for general three-momentum we use spherical polar
coordinates

\begin{equation}
\hat{\boldsymbol{p}}=\left(\sin\theta\cos\varphi,\sin\theta\sin\varphi,\cos\theta\right),\label{eq:2.7}
\end{equation}

\noindent and the transformation

\begin{equation}
\Lambda\left(\theta,\varphi\right)=\exp\left\{ -\dfrac{\theta}{2}\left(\gamma^{1}\cos\varphi+\gamma^{2}\sin\varphi\right)\gamma^{3}\right\} ,\label{eq:2.8}
\end{equation}

\noindent which is actually a rotation since it is unitary and of
unit determinant. Applying Eq. (\ref{eq:2.8}) to the bispinors in
Eq. (\ref{eq:2.6}) we have

\begin{equation}
\begin{array}{ccc}
\Lambda\left(\theta,\varphi\right)u^{(i)}\left(p_{z}\right)=u^{(i)}(\mathbf{p}), &  & i=1,\ldots,4,\end{array}\label{eq:2.9}
\end{equation}

\noindent with the general momentum bispinors, in two-block notation,
given by

\begin{equation}
\begin{array}{cc}
u^{(1)}(\mathbf{p})=\begin{pmatrix}0\\
\chi_{+}\left(\mathbf{p}\right)
\end{pmatrix}, & u^{(2)}(\mathbf{p})=\begin{pmatrix}\chi_{-}\left(\mathbf{p}\right)\\
0
\end{pmatrix},\\
\\
u^{(3)}(\mathbf{p})=\begin{pmatrix}0\\
\chi_{-}\left(\mathbf{p}\right)
\end{pmatrix}, & u^{(4)}(\mathbf{p})=\begin{pmatrix}\chi_{+}\left(\mathbf{p}\right)\\
0
\end{pmatrix},
\end{array}\label{eq:2.10}
\end{equation}

\noindent where $\chi_{\pm}\left(\mathbf{p}\right)$ are the two-component
helicity eigenspinors

\begin{align}
\begin{split}\chi_{+}\left(\mathbf{p}\right)= & \begin{pmatrix}\cos\left(\frac{\theta}{2}\right)\\
e^{i\varphi}\sin\left(\frac{\theta}{2}\right)
\end{pmatrix},\\
\chi_{-}\left(\mathbf{p}\right)= & \begin{pmatrix}-e^{-i\varphi}\sin\left(\frac{\theta}{2}\right)\\
\cos\left(\frac{\theta}{2}\right)
\end{pmatrix},
\end{split}
\label{eq:2.11}
\end{align}

\noindent satisfying the equation

\begin{equation}
\boldsymbol{\sigma}\cdot\mathbf{\hat{p}}\,\chi_{\pm}(\mathbf{p})=\pm\chi_{\pm}(\mathbf{p}).\label{eq:2.12}
\end{equation}

The bispinors in Eq. (\ref{eq:2.10}) are orthonormal

\begin{equation}
u^{\dagger(i)}(\mathbf{p})u^{(j)}(\mathbf{p})=\delta_{ij},\label{eq:2.12-1}
\end{equation}

\noindent with a normalization that is adequate for massless spinors,
as the Dirac adjoint $\overline{u}\equiv u^{\dagger}\gamma^{0}$ is
not needed in this case. Another useful, Lorentz invariant normalization
is to re-scale them to $\sqrt{2E}$. These bispinors are also solutions
to Eq. (\ref{eq:2.5}), which in Hamiltonian form reads

\begin{equation}
\begin{split}\boldsymbol{\alpha}\cdot\hat{\mathbf{p}}\,u^{(s)}(\mathbf{p})= & +u^{(s)}(\mathbf{p}),\\
\boldsymbol{\alpha}\cdot\hat{\mathbf{p}}\,u^{(s+2)}(\mathbf{p})= & -u^{(s+2)}(\mathbf{p}),
\end{split}
s=1,2\label{eq:2.13}
\end{equation}

\noindent making explicit that $u^{(1)}(\mathbf{p})$ and $u^{(2)}(\mathbf{p})$
are positive-energy bispinors, while $u^{(3)}(\mathbf{p})$ and $u^{(4)}(\mathbf{p})$
are negative-energy ones. The helicity and chirality eigenvalues are
the same as in Eq. (\ref{eq:2.6}). Energy projection operators are
obtained from the spin sums

\begin{align}
\begin{split}\varLambda_{+}\equiv & {\displaystyle \sum_{s=1,2}}u^{(s)}(\mathbf{p})u^{\dagger(s)}(\mathbf{p})=\frac{1}{2}\left(\mathbbm{1}+\boldsymbol{\alpha}\cdot\hat{\mathbf{p}}\right),\\
\varLambda_{-}\equiv & {\displaystyle \sum_{s=1,2}}u^{(s+2)}(\mathbf{p})u^{\dagger(s+2)}(\mathbf{p})=\frac{1}{2}\left(\mathbbm{1}-\boldsymbol{\alpha}\cdot\hat{\mathbf{p}}\right).
\end{split}
\label{eq:2.14}
\end{align}

\noindent They satisfy the required properties for projection operators

\begin{gather}
\begin{gathered}\varLambda_{\pm}^{2}=\varLambda_{\pm},\\
\varLambda_{+}\varLambda_{-}=\varLambda_{-}\varLambda_{+}=0,\\
\varLambda_{+}+\varLambda_{-}=1,
\end{gathered}
\label{eq:2.15}
\end{gather}

\noindent and from the second and third properties it is readily seen
that the bispinors in Eq. (\ref{eq:2.10}) constitute a complete and
orthogonal set of solutions to the massless Dirac equation. The energy
projection operators in Eq. (\ref{eq:2.14}) can be found in the literature\citep{Malekzadeh:2008im,PhysRevD.61.051501,Pisarski:1999tv,Reuter:2006rf,PhysRevD.76.125022,Rho:2000ww},
although without reference to the bispinors and the spin sums.

\subsection{Majorana}

Using the canonical frame bispinors in Eq. (\ref{eq:2.6}) we define
the following Majorana bispinors

\begin{align}
\begin{split}u_{M}^{(1)}\left(p_{z}\right)= & \dfrac{1}{\sqrt{2}}\left(u^{(2)}\left(p_{z}\right)+i\gamma^{2}u^{*(2)}\left(p_{z}\right)\right),\\
u_{M}^{(2)}\left(p_{z}\right)= & \dfrac{1}{\sqrt{2}}\left(u^{(1)}\left(p_{z}\right)-i\gamma^{2}u^{*(1)}\left(p_{z}\right)\right),\\
u_{M}^{(3)}\left(p_{z}\right)= & \dfrac{1}{\sqrt{2}}\left(u^{(3)}\left(p_{z}\right)-i\gamma^{2}u^{*(3)}\left(p_{z}\right)\right),\\
u_{M}^{(4)}\left(p_{z}\right)= & \dfrac{1}{\sqrt{2}}\left(u^{(4)}\left(p_{z}\right)+i\gamma^{2}u^{*(4)}\left(p_{z}\right)\right),
\end{split}
\label{eq:2.16}
\end{align}

\noindent where the asterisk denotes complex conjugation, even though
it is superfluous in this case because the $u^{(i)}\left(p_{z}\right)$
are real. The bispinors in Eq. (\ref{eq:2.16}) are eigenstates of
the standard charge conjugation operator\citep{halzen1984quarks,peskin1995introduction} 

\begin{equation}
\mathcal{C}\equiv CK\equiv i\gamma^{2}K,\label{eq:2.17}
\end{equation}

\noindent where $C=i\gamma^{2}$ is the charge conjugation matrix,
and $K$ stands for the operation of complex conjugation to the right.
We then have

\begin{align}
\begin{split}\mathcal{C}u_{M}^{(1,4)}\left(p_{z}\right)= & +u_{M}^{(1,4)}\left(p_{z}\right),\\
\mathcal{C}u_{M}^{(2,3)}\left(p_{z}\right)= & -u_{M}^{(2,3)}\left(p_{z}\right),
\end{split}
\label{eq:2.18}
\end{align}

\noindent and it is in this sense that they fulfill the Majorana condition.
These Majorana bispinors are also solutions to Eq. (\ref{eq:2.13}),
implying a unitary transformation must exist relating them to the
Weyl bispinors in Eq. (\ref{eq:2.6}). Among several possibilities,
to be discussed in the next section, we choose

\begin{equation}
R_{3}=\exp\left(\frac{\pi}{4}\gamma^{0}\gamma^{1}\gamma^{3}\right),\label{eq:2.19}
\end{equation}

\noindent as the transformation matrix, which besides being unitary
is also of unit determinant, therefore a rotation. Thus, we have the
following equivalence between the bispinors in Eqs. (\ref{eq:2.6})
and (\ref{eq:2.16})

\begin{equation}
\begin{array}{cc}
R_{3}u^{(1)}\left(p_{z}\right)=-u_{M}^{(1)}\left(p_{z}\right), & R_{3}u^{(2)}\left(p_{z}\right)=+u_{M}^{(2)}\left(p_{z}\right),\\
R_{3}u^{(3)}\left(p_{z}\right)=+u_{M}^{(4)}\left(p_{z}\right), & R_{3}u^{(4)}\left(p_{z}\right)=-u_{M}^{(3)}\left(p_{z}\right).
\end{array}\label{eq:2.20}
\end{equation}

It is now straightforward to generalize this result to arbitrary momentum
bispinors. Using the ones in Eq. (\ref{eq:2.10}) we obtain the generalization
of Eq. (\ref{eq:2.16})

\begin{equation}
\begin{split}u_{M}^{(1)}\left(\boldsymbol{p}\right)= & \dfrac{1}{\sqrt{2}}\left(u^{(2)}\left(\boldsymbol{p}\right)+i\gamma^{2}u^{(2)^{*}}\left(\boldsymbol{p}\right)\right),\\
u_{M}^{(2)}\left(\boldsymbol{p}\right)= & \dfrac{1}{\sqrt{2}}\left(u^{(1)}\left(\boldsymbol{p}\right)-i\gamma^{2}u^{(1)^{*}}\left(\boldsymbol{p}\right)\right),\\
u_{M}^{(3)}\left(\boldsymbol{p}\right)= & \dfrac{1}{\sqrt{2}}\left(u^{(3)}\left(\boldsymbol{p}\right)-i\gamma^{2}u^{(3)^{*}}\left(\boldsymbol{p}\right)\right),\\
u_{M}^{(4)}\left(\boldsymbol{p}\right)= & \dfrac{1}{\sqrt{2}}\left(u^{(4)}\left(\boldsymbol{p}\right)+i\gamma^{2}u^{(4)^{*}}\left(\boldsymbol{p}\right)\right).
\end{split}
\label{eq:2.21}
\end{equation}

\noindent These Majorana bispinors are obtained from the canonical
frame ones in Eq. (\ref{eq:2.16}) by the same rotation in Eq. (\ref{eq:2.8})

\begin{equation}
\begin{array}{ccc}
\Lambda\left(\theta,\varphi\right)u_{M}^{(i)}\left(p_{z}\right)=u_{M}^{(i)}(\mathbf{p}), &  & i=1,\ldots,4.\end{array}\label{eq:2.21-1}
\end{equation}

\noindent Then defining the rotation

\noindent 
\begin{equation}
\Omega\left(\theta,\varphi\right)\equiv\Lambda\left(\theta,\varphi\right)R_{3}\Lambda^{\dagger}\left(\theta,\varphi\right),\label{eq:2.21-2}
\end{equation}

\noindent equations (\ref{eq:2.9}) and (\ref{eq:2.20}) yield

\begin{equation}
\begin{array}{cc}
\Omega\left(\theta,\varphi\right)u^{(1)}\left(\boldsymbol{p}\right)=-u_{M}^{(1)}\left(\boldsymbol{p}\right), & \Omega\left(\theta,\varphi\right)u^{(2)}\left(\boldsymbol{p}\right)=+u_{M}^{(2)}\left(\boldsymbol{p}\right),\\
\Omega\left(\theta,\varphi\right)u^{(3)}\left(\boldsymbol{p}\right)=+u_{M}^{(4)}\left(\boldsymbol{p}\right), & \Omega\left(\theta,\varphi\right)u^{(4)}\left(\boldsymbol{p}\right)=-u_{M}^{(3)}\left(\boldsymbol{p}\right).
\end{array}\label{eq:2.22}
\end{equation}

\noindent Observing that $\Omega\left(\theta,\varphi\right)$ and
$\boldsymbol{\alpha}\cdot\hat{\mathbf{p}}$ commute, it is readily
verified that the bispinors in Eq. (\ref{eq:2.21}) are solutions
to the massless Dirac equation

\begin{equation}
\begin{split}\boldsymbol{\alpha}\cdot\hat{\mathbf{p}}\,u_{M}^{(s)}(\mathbf{p})= & +u_{M}^{(s)}(\mathbf{p}),\\
\boldsymbol{\alpha}\cdot\hat{\mathbf{p}}\,u_{M}^{(s+2)}(\mathbf{p})= & -u_{M}^{(s+2)}(\mathbf{p}).
\end{split}
s=1,2,\label{eq:2.23}
\end{equation}

\noindent They also satisfy the Majorana condition 

\begin{equation}
\begin{split}\mathcal{C}u_{M}^{(1,4)}\left(\boldsymbol{p}\right)= & +u_{M}^{(1,4)}\left(\boldsymbol{p}\right),\\
\mathcal{C}u_{M}^{(2,3)}\left(\boldsymbol{p}\right)= & -u_{M}^{(2,3)}\left(\boldsymbol{p}\right).
\end{split}
\label{eq:2.24}
\end{equation}

\noindent Accordingly, Eq. (\ref{eq:2.22}) establishes an equivalence
between Weyl and massless Majorana bispinors. This relation is the
$c$-number analogue of the known equivalence between Weyl and massless
Majorana \emph{field operators},\emph{ }related by a Pauli-Gursey
transformation.\citep{Zralek:1997sa,fukugita2003physics,boyarkin2011advanced}
In this sense this result completes the equivalence between massless
Majorana and Weyl fermions, which is now seen to hold for both quantum
fields and $c$-number spinors.

\subsection{Completeness and degrees of freedom}

There is a subtle but important matter regarding the negative-energy
Weyl bispinors $u^{(3,4)}(\mathbf{p})$ in Eq. (\ref{eq:2.10}). If
one substitutes the complete wavefunctions $\Psi^{(3,4)}(x)=u^{(3,4)}(\mathbf{p})e^{ip.x}$
in Eq. (\ref{eq:2.1}) it is found that $\boldsymbol{\alpha}\cdot\hat{\mathbf{p}}\,u^{(3,4)}(\mathbf{p})=u^{(3,4)}(\mathbf{p})$,
in contradiction with Eq. (\ref{eq:2.13}). Let us contrast this situation
with the standard massive case\citep{halzen1984quarks,aitchison2002gauge}
where, following the Feynman - Stuckelberg prescription for antiparticles,
the negative-energy bispinors are redefined as $v_{m}^{(1,2)}(\mathbf{p})\equiv u_{m}^{(4,3)}(-\mathbf{p})$
(the subscript $m$ is just to make explicit that these are massive
bispinors). The momentum flip is necessary so that solutions with
four-momentum $(-E,-\mathbf{p})$ are interpreted ($E$ is always
positive) as antiparticle solutions with four-momentum $(E,\mathbf{p})$,
and the coordinate dependence is obtained from the positive-energy
one $e^{-ip.x}$ by making the replacements $E\rightarrow-E$ and
$\mathbf{p}\rightarrow-\mathbf{p}$. Also, the spinors indexes are
relabeled to implement hole theory in the rest frame. 

In the massless case there is no rest frame, but one can use the canonical
frame instead, with helicity replacing spin in hole theory. Hence,
the absence of a negative-energy solution with positive (negative)
helicity, and therefore negative (positive) chirality, is to be interpreted
as the presence of a positive energy solution with negative (positive)
helicity, and the same chirality. The momentum flip is still necessary
for the antiparticle interpretation, and in fact it is already implied
for the plane wave $e^{ip.x}$, with $p^{0}=E\equiv\left|\mathbf{p}\right|$,
but combined with a simple relabeling of the spinor indexes, as in
the massive case, is not enough to satisfy helicity invariance. Hence,
both spin and momentum of the negative-energy solutions must be reversed.
However these operations just produce the same bispinors up to a phase.
To see it, it suffices to consider the spinors in Eq. (\ref{eq:2.11}).
The momentum flip is accomplished through the substitution $\left(\theta,\varphi\right)\rightarrow\left(\pi-\theta,\phi+\pi\right)$,
leading to

\begin{equation}
\chi_{\pm}(-\mathbf{p})=\mp e^{\pm i\varphi}\chi_{\mp}(\mathbf{p}),\label{eq:3.1}
\end{equation}

\noindent while the spin flip is done via\citep{peskin1995introduction}

\begin{equation}
-i\sigma_{2}\chi_{\pm}^{*}(\mathbf{p})=\pm\chi_{\mp}(\mathbf{p}).\label{eq:3.2}
\end{equation}

\noindent Thus, Eqs. (\ref{eq:3.1}) and (\ref{eq:3.2}) produce

\begin{equation}
-i\sigma_{2}\chi_{\pm}^{*}(-\mathbf{p})=e^{\mp i\varphi}\chi_{\pm}(\mathbf{p}).\label{eq:3.3}
\end{equation}

\noindent As for the coordinate dependence, and starting from $\exp\left\{ -i\left(-Et-\boldsymbol{p}\cdot\boldsymbol{x}\right)\right\} $,
the operations of complex conjugating and flipping the momentum result
in the positive-energy case $e^{-ip.x}$.

In the literature, the above discrepancy is expressed in terms of
incompatible statements about the completeness of solutions to the
massless Dirac equation. On one hand, in Refs. \citep{Lurie1968,Dixon:1996wi,gross1999relativistic,itzykson2006quantum,srednicki2007quantum,schwabl2008advanced,das2008lectures}
it is concluded, following different approaches, that there are only
two independent solutions to the equation, with the negative-energy
bispinors being proportional to the positive-energy ones. On the other
hand, the massless Dirac equation in momentum space is a $4\times4$
Hermitian matrix, so there must be four independent solutions, as
already given in Eqs. (\ref{eq:2.6}) and (\ref{eq:2.10}), and expressed
in the completeness relations in Eq. (\ref{eq:2.14}). The resolution
of this problem lies in the degrees of freedom: A Majorana bispinor,
either massless or massive, possesses two degrees of freedom because
of the Majorana condition, and these are half the degrees of freedom
of a Dirac bispinor. In view of the results of the last subsection,
this is also true for the Weyl bispinors. Thus, even if formally four
independent solutions exist for the massless Dirac equation (a complete
set in the mathematical sense), only two make sense physically. 

At the level of $c$-number wave functions one could, in principle,
either give up the Feynman - Stuckelberg interpretation for negative-energy
states and keep the complete set of solutions, or maintain the conceptually
useful antiparticle interpretation and disregard mathematical completeness,
since ultimately it is the quantized theory (second quantization)
the one that is expected to be free of ambiguities. Indeed, in a classic
paper\citep{PhysRev.134.B882} Weinberg has shown that, under the
general assumption of Lorentz invariance of the $S$ matrix, massless
fermionic field operators must be given by

\begin{align}
\begin{split}\psi_{-}(x)= & \int\frac{d^{3}p}{(2\pi)^{3}\sqrt{2E}}\left(a_{-}\left(\boldsymbol{p}\right)e^{-ip\cdot x}+b_{+}^{\dagger}\left(\boldsymbol{p}\right)e^{ip\cdot x}\right)\sqrt{2E}\chi_{-}\left(\mathbf{p}\right),\\
\psi_{+}(x)= & \int\frac{d^{3}p}{(2\pi)^{3}\sqrt{2E}}\left(a_{+}\left(\boldsymbol{p}\right)e^{-ip\cdot x}+b_{-}^{\dagger}\left(\boldsymbol{p}\right)e^{ip\cdot x}\right)\sqrt{2E}\chi_{+}\left(\mathbf{p}\right),
\end{split}
\label{eq:3.4}
\end{align}

\noindent where the subscripts $\pm$ respectively represent positive
and negative helicity, and the spinors $\chi_{\pm}\left(\mathbf{p}\right)$
are given in Eq. (\ref{eq:2.11}). These massless fields can be readily
expressed in terms of bispinors by making the substitutions $\chi_{\pm}\left(\mathbf{p}\right)\rightarrow u^{(1,2)}(\mathbf{p})$,
with the latter given in Eq. (\ref{eq:2.10}). There is no use for
the complete set of massless bispinors in the field operator expansion.

\section{Majorana condition and maximal entanglement}

\subsection{Massless bispinors as bipartite qubits}

In quantum computation the quantum analogue of a classical bit, a
qubit, is given by a complex linear combination of the basis states
of a two-level quantum system, known as the computational basis. Denoting
the basis states by $\left|0\right\rangle $ and $\left|1\right\rangle $,
for spin-$1/2$ systems they can be chosen as the eigenstates of $\sigma^{3}$

\begin{equation}
\begin{array}{cc}
\left|0\right\rangle =\left(\begin{array}{c}
1\\
0
\end{array}\right), & \left|1\right\rangle =\left(\begin{array}{c}
0\\
1
\end{array}\right).\end{array}\label{eq:4.1}
\end{equation}

\noindent In this basis, the helicity spinors in Eq. (\ref{eq:2.11})
are given by the general pure-state qubits

\begin{align}
\begin{split}\left|\chi_{+}\right\rangle = & \cos\left(\frac{\theta}{2}\right)\left|0\right\rangle +e^{i\varphi}\sin\left(\frac{\theta}{2}\right)\left|1\right\rangle ,\\
\left|\chi_{-}\right\rangle = & -e^{-i\varphi}\sin\left(\frac{\theta}{2}\right)\left|0\right\rangle +\cos\left(\frac{\theta}{2}\right)\left|1\right\rangle ,
\end{split}
\label{eq:4.2}
\end{align}

\noindent which are antipodal in the unit Bloch sphere representation,\citep{barnett2009quantum,nielsen2010quantum}
with the three-momentum in Eq. (\ref{eq:2.7}) taken as the Bloch
vector (Fig. 1).

\begin{figure}
\begin{centering}
\includegraphics[scale=0.4]{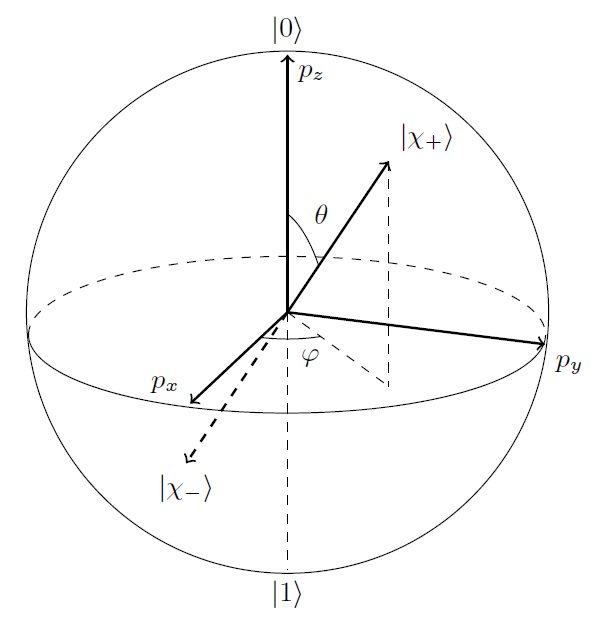}
\par\end{centering}
\caption{Unit Bloch Sphere. The computational basis is mapped to the north
and south poles of the sphere. The orthogonal pure states $\left|\chi_{+}\right\rangle $
and $\left|\chi_{-}\right\rangle $ are antipodal and correspond to
helicity eigenspinors if the Bloch vector is taken as the three-momentum.}
\end{figure}

The computational basis for the space of two pure-state qubits is
then given by the set $\left\{ \left|0\right\rangle ,\left|1\right\rangle \right\} \otimes\left\{ \left|0\right\rangle ,\left|1\right\rangle \right\} $,
whence, upon using Eq. (\ref{eq:4.1}) and the notation $\left|00\right\rangle \equiv\left|0\right\rangle \otimes\left|0\right\rangle $
and so on, we obtain the explicit representation

\begin{equation}
\begin{array}{cccc}
\left|00\right\rangle =\begin{pmatrix}1\\
0\\
0\\
0
\end{pmatrix}, & \left|01\right\rangle =\begin{pmatrix}0\\
1\\
0\\
0
\end{pmatrix}, & \left|10\right\rangle =\begin{pmatrix}0\\
0\\
1\\
0
\end{pmatrix}, & \left|11\right\rangle =\begin{pmatrix}0\\
0\\
0\\
1
\end{pmatrix}\end{array},\label{eq:4.3}
\end{equation}

\noindent and we see that the elements of the basis are just the canonical
frame Weyl bispinors in Eq. (\ref{eq:2.6})

\begin{equation}
\begin{array}{cc}
\left|00\right\rangle =u^{(4)}\left(p_{z}\right), & \left|01\right\rangle =u^{(2)}\left(p_{z}\right),\\
\left|10\right\rangle =u^{(1)}\left(p_{z}\right), & \left|11\right\rangle =u^{(3)}\left(p_{z}\right).
\end{array}\label{eq:4.4}
\end{equation}

\noindent Another basis for this space is provided by the Bell states,
which are maximally entangled states

\begin{align}
\begin{split}\left|\Phi^{+}\right\rangle = & \dfrac{1}{\sqrt{2}}\left(\left|00\right\rangle +\left|11\right\rangle \right),\\
\left|\Phi^{-}\right\rangle = & \dfrac{1}{\sqrt{2}}\left(\left|00\right\rangle -\left|11\right\rangle \right),\\
\left|\Psi^{+}\right\rangle = & \dfrac{1}{\sqrt{2}}\left(\left|01\right\rangle +\left|10\right\rangle \right),\\
\left|\Psi^{-}\right\rangle = & \dfrac{1}{\sqrt{2}}\left(\left|01\right\rangle -\left|10\right\rangle \right).
\end{split}
\label{eq:4.5}
\end{align}

\noindent Using either of Eqs. (\ref{eq:2.6}) or (\ref{eq:4.3}),
explicit representations of the Bell states, as well as the massless
Majorana bispinors in Eq. (\ref{eq:2.16}), are directly obtained,
and upon comparing the two sets we arrive at the interesting result
that the Bell states are algebraically equivalent to the massless
Majorana bispinors in the canonical frame 

\begin{equation}
\begin{array}{lc}
u_{M}^{(1)}\left(p_{z}\right)=\left|\Psi^{-}\right\rangle , & u_{M}^{(2)}\left(p_{z}\right)=\left|\Psi^{+}\right\rangle ,\\
u_{M}^{(3)}\left(p_{z}\right)=-\left|\Phi^{-}\right\rangle , & u_{M}^{(4)}\left(p_{z}\right)=\left|\Phi^{+}\right\rangle .
\end{array}\label{eq:4.6}
\end{equation}
This result is generalized to arbitrary momentum defining the general-momentum
Bell states

\begin{equation}
\begin{split}\left|\Phi^{+}(\mathbf{p})\right\rangle = & \dfrac{1}{\sqrt{2}}\left(u^{(4)}(\mathbf{p})+u^{(3)}(\mathbf{p})\right),\\
\left|\Phi^{-}(\mathbf{p})\right\rangle = & \dfrac{1}{\sqrt{2}}\left(u^{(4)}(\mathbf{p})-u^{(3)}(\mathbf{p})\right),\\
\left|\Psi^{+}(\mathbf{p})\right\rangle = & \dfrac{1}{\sqrt{2}}\left(u^{(2)}(\mathbf{p})+u^{(1)}(\mathbf{p})\right),\\
\left|\Psi^{-}(\mathbf{p})\right\rangle = & \dfrac{1}{\sqrt{2}}\left(u^{(2)}(\mathbf{p})-u^{(1)}(\mathbf{p})\right),
\end{split}
\label{eq:4.7}
\end{equation}

\noindent then, from Eqs. (\ref{eq:2.10}), (\ref{eq:2.11}), and
(\ref{eq:2.21}) we get

\begin{equation}
\begin{array}{lc}
u_{M}^{(1)}(\mathbf{p})=\left|\Psi^{-}(\mathbf{p})\right\rangle , & u_{M}^{(2)}(\mathbf{p})=\left|\Psi^{+}(\mathbf{p})\right\rangle ,\\
u_{M}^{(3)}(\mathbf{p})=-\left|\Phi^{-}(\mathbf{p})\right\rangle , & u_{M}^{(4)}(\mathbf{p})=\left|\Phi^{+}(\mathbf{p})\right\rangle .
\end{array}\label{eq:4.8}
\end{equation}

\noindent Thus, we conclude that for massless bispinors obeying the
Dirac equation, the Majorana condition is equivalent to maximal entanglement.

\subsection{Entangling gates}

Operations on qubits are given by unitary quantum gates, and from
Eqs. (\ref{eq:2.20}) and (\ref{eq:4.6}) we see that the rotation
in Eq (\ref{eq:2.19}) serves as a two-qubit gate that produces entanglement.
We now provide a list, not meant to be exhaustive, of other entangling
gates and their properties.

The common procedure for producing entanglement in quantum computation
is by a combination of a CNOT (controlled not) gate and a Hadamard
gate. The latter is a one-qubit gate given by

\begin{align}
\begin{split}H= & \dfrac{1}{\sqrt{2}}\left(\left|0\right\rangle \left\langle 0\right|+\left|1\right\rangle \left\langle 0\right|+\left|0\right\rangle \left\langle 1\right|-\left|1\right\rangle \left\langle 1\right|\right)\\
= & \dfrac{1}{\sqrt{2}}\begin{pmatrix}1 & 1\\
1 & -1
\end{pmatrix},
\end{split}
\label{eq:4.9}
\end{align}

\noindent while the former is a two-qubit gate, with the most common
realization given by

\begin{equation}
C_{NOT}=\begin{pmatrix}1 & 0 & 0 & 0\\
0 & 1 & 0 & 0\\
0 & 0 & 0 & 1\\
0 & 0 & 1 & 0
\end{pmatrix}.\label{eq:4.10}
\end{equation}

\noindent Then it is easy to see that, say, the combination $C_{NOT}\left(H\otimes\mathbbm{1}_{2}\right)$
produces the Bell states in Eq. (\ref{eq:4.5}) when acting on the
computational basis in Eq. (\ref{eq:4.3}), e.g, $C_{NOT}\left(H\otimes\mathbbm{1}_{2}\right)\left|00\right\rangle =\left|\Phi^{+}\right\rangle $.
The CNOT is a universal gate\citep{nielsen2010quantum} in the sense
that any quantum circuit can be simulated with arbitrary accuracy
by a combination of a CNOT and one-qubit gates (the latter usually
taken as the Hadamard and the $\pi/8$ phase gates). It has also been
shown, for the two-qubit case, that the relevant property for universality
is entanglement,\citep{Brylinski2002} and so any quantum circuit
can be simulated with arbitrary accuracy by a combination of an entangling
two-qubit gate and suitable one-qubit gates. It is also worth noticing
that the CNOT gate is not a rotation, since it has determinant -1,
a feature that difficults actual implementations.

Another set of entangling gates, denoted by $R_{i}$, $i=1,\ldots,4$,
consists of the rotations 

\begin{align}
\begin{split}R_{1}= & \exp\left(\frac{\pi}{4}\gamma^{1}\right),\\
R_{2}= & \exp\left(-\frac{\pi}{4}\gamma^{1}\right),\\
R_{3}= & \exp\left(\frac{\pi}{4}\gamma^{0}\gamma^{1}\gamma^{3}\right),\\
R_{4}= & \exp\left(-\frac{\pi}{4}\gamma^{0}\gamma^{1}\gamma^{3}\right).
\end{split}
\label{eq:4.11}
\end{align}

\noindent They also have the interesting property of being solutions
to the algebraic Yang-Baxter equation.\citep{2006math.ph...6053P}

\begin{equation}
\left(R_{i}\otimes\mathbbm{1}_{2}\right)\left(\mathbbm{1}_{2}\otimes R_{i}\right)\left(R_{i}\otimes\mathbbm{1}_{2}\right)=\left(\mathbbm{1}_{2}\otimes R_{i}\right)\left(R_{i}\otimes\mathbbm{1}_{2}\right)\left(\mathbbm{1}_{2}\otimes R_{i}\right).\label{eq:4.12}
\end{equation}

\noindent These matrices have been studied by Kauffman et al\citep{Kauffman2019}
in connection with knot theory and topological linking. The gate $R_{3}$
(used in Eq. (\ref{eq:2.21-2})) was introduced by Kauffman and Lomonaco\citep{1367-2630-6-1-134},
while the matrices $R_{1}$ and $R_{2}$ appear, respectively, in
Refs. \onlinecite{1751-8121-41-5-055310} and \onlinecite{0295-5075-92-3-30002}.
The action of these gates on the computational basis is summarized
in Table 2.

\begin{table}
\begin{centering}
\begin{tabular}{c|r|r|r|r|}
\hline 
 & $\left|10\right\rangle $ & $\left|01\right\rangle $ & $\left|11\right\rangle $ & $\left|00\right\rangle $\tabularnewline
\hline 
\hline 
$R_{1}$ & $\left|\Psi^{+}\right\rangle $ & $\left|\Psi^{-}\right\rangle $ & $\left|\Phi^{+}\right\rangle $ & $\left|\Phi^{-}\right\rangle $\tabularnewline
\hline 
$R_{2}$ & $-\left|\Psi^{-}\right\rangle $ & $\left|\Psi^{+}\right\rangle $ & $-\left|\Phi^{-}\right\rangle $ & $\left|\Phi^{+}\right\rangle $\tabularnewline
\hline 
$R_{3}$ & $-\left|\Psi^{-}\right\rangle $ & $\left|\Psi^{+}\right\rangle $ & $\left|\Phi^{+}\right\rangle $ & $\left|\Phi^{-}\right\rangle $\tabularnewline
\hline 
$R_{4}$ & $\left|\Psi^{+}\right\rangle $ & $\left|\Psi^{-}\right\rangle $ & $-\left|\Phi^{-}\right\rangle $ & $\left|\Phi^{+}\right\rangle $\tabularnewline
\hline 
\end{tabular}
\par\end{centering}
\caption{Action of the entangling gates in Eq. (\ref{eq:4.11}) on the computational
basis in Eq. (\ref{eq:4.3}). The table is read so that the gates
in the first column act on the basis states in the top first row and
produce the given Bell state in the intersection.}
\end{table}

Yet another set of entangling gates, denoted by $\hat{R}_{i}$, $i=1,\ldots,4$,
is given by the rotations

\begin{align}
\begin{split}\hat{R}_{1}= & \frac{i}{\sqrt{2}}\gamma^{3}\left(\mathbbm{1}+\gamma^{1}\right),\\
\hat{R}_{2}= & \frac{i}{\sqrt{2}}\gamma^{2}\left(\mathbbm{1}+\gamma^{1}\right),\\
\hat{R}_{3}= & \frac{1}{\sqrt{2}}\gamma^{0}\left(\mathbbm{1}+\gamma^{1}\right),\\
\hat{R}_{4}= & \frac{i}{\sqrt{2}}\left(\gamma^{0}\gamma^{2}\gamma^{3}+i\gamma^{5}\right).
\end{split}
\label{eq:4.13}
\end{align}

\noindent They are also Hermitian and therefore square to the identity
matrix

\begin{equation}
\begin{array}{ccc}
\hat{R}_{i}=\hat{R}_{i}^{\dagger}, & \hat{R}_{i}^{2}=\mathbbm{1}, & i=1,\ldots,4.\end{array}\label{eq:4.14}
\end{equation}

\noindent They do not obey Eq. (\ref{eq:4.12}), but instead satisfy
the anti-commutation (Clifford algebra) relations

\begin{equation}
\left\{ \hat{R}_{i},\hat{R}_{j}^{\dagger}\right\} =2\delta_{i,j}\mathbbm{1},\label{eq:4.15}
\end{equation}

\noindent which, in contrast, are not obeyed by the gates in Eq. (\ref{eq:4.11}).
These matrices are all orthogonal to each other, as is verified with
the inner product

\noindent 
\begin{equation}
\text{Tr}\left(\hat{R}_{i}^{\dagger}\hat{R}_{j}\right)=0,\,\,\,i,j=1,\ldots,4,\,\,\,i\neq j,\label{eq:4.16}
\end{equation}

\noindent hence, they are linearly independent. Using the 16 elements
of the 4-d gamma matrices Clifford algebra it can be verified that
no other matrix exists with these characteristics that fulfils Eq.
(\ref{eq:4.14}) and also closes the algebra in Eq. (\ref{eq:4.15}).
In this sense the set in Eq. (\ref{eq:4.13}) is complete. Their action
on the computational basis is shown in Table 3.

\begin{table}
\begin{centering}
\begin{tabular}{c|r|r|r|r|}
\hline 
 & $\left|10\right\rangle $ & $\left|01\right\rangle $ & $\left|11\right\rangle $ & $\left|00\right\rangle $\tabularnewline
\hline 
\hline 
$\hat{R}_{1}$ & $i\left|\Phi^{+}\right\rangle $ & $-i\left|\Phi^{-}\right\rangle $ & $-i\left|\Psi^{+}\right\rangle $ & $i\left|\Psi^{-}\right\rangle $\tabularnewline
\hline 
$\hat{R}_{2}$ & $-\left|\Psi^{+}\right\rangle $ & $\left|\Psi^{-}\right\rangle $ & $\left|\Phi^{+}\right\rangle $ & $-\left|\Phi^{-}\right\rangle $\tabularnewline
\hline 
$\hat{R}_{3}$ & $\left|\Phi^{+}\right\rangle $ & $-\left|\Phi^{-}\right\rangle $ & $\left|\Psi^{+}\right\rangle $ & $-\left|\Psi^{-}\right\rangle $\tabularnewline
\hline 
$\hat{R}_{4}$ & $\left|\Psi^{-}\right\rangle $ & $\left|\Psi^{+}\right\rangle $ & $\left|\Phi^{-}\right\rangle $ & $\left|\Phi^{+}\right\rangle $\tabularnewline
\hline 
\end{tabular}
\par\end{centering}
\caption{Action of the entangling gates in Eq. (\ref{eq:4.13}) on the computational
basis in Eq. (\ref{eq:4.3}). The table is read so that the gates
in the first column act on the basis states in the top first row and
produce the given Bell state in the intersection.}
\end{table}

\section{Quantum computational toy model with four Majorana zero modes}

The properties in Eqs. (\ref{eq:4.14}) and (\ref{eq:4.15}), obeyed
by the $\hat{R}_{i}$ gates, are the same as the ones satisfied by
Majorana zero mode operators\citep{0268-1242-27-12-124003,doi:10.1146/annurev-conmatphys-030212-184337,Sarma2015}
in topological quantum computation. Hence, we will regard them as
such and present a general model with four Majorana bound states that
admits entanglement from braiding. We mostly follow Ref. \onlinecite{PhysRevA.73.042313}.

The set $\mathcal{M}$ of particle types is given by

\begin{equation}
\mathcal{M}=\left\{ \boldsymbol{1}_{\text{vac}},\sigma,\psi\right\} ,\label{eq:5.1}
\end{equation}

\noindent consisting of the vacuum $\boldsymbol{1}_{\text{vac}}$,
anyons $\sigma$, and Weyl fermions $\psi$, with the standard fusion
rules

\begin{gather}
\begin{gathered}g\times\boldsymbol{1}_{\text{vac}}=g,\,\,\,\forall g\in\mathcal{M},\\
\sigma\times\psi=\sigma,\\
\psi\times\psi=\boldsymbol{1}_{\text{vac}},\\
\sigma\times\sigma=\boldsymbol{1}_{\text{vac}}+\psi.
\end{gathered}
\label{eq:5.2}
\end{gather}

To define braid operators we take the branch cuts of the Majorana
zero modes, described by the Majorana operators, in the same direction,
and order them in a way that exchanging $\hat{R}_{i}$ and $\hat{R}_{i+1}$
clockwise ensures that $\hat{R}_{i}$ crosses solely the branch cut
of $\hat{R}_{i+1}$, with no other operator crossing any other branch
cut. Then the local (nearest-neighbor) braid operators are given by 

\begin{align}
\begin{split}B_{12}= & \exp\left(-\frac{\pi}{4}\hat{R}_{1}\hat{R}_{2}\right),\\
B_{23}= & \exp\left(-\frac{\pi}{4}\hat{R}_{2}\hat{R}_{3}\right),\\
B_{34}= & \exp\left(-\frac{\pi}{4}\hat{R}_{3}\hat{R}_{4}\right).
\end{split}
\label{eq:5.3}
\end{align}

\noindent They are unitary by construction, and satisfy the required
properties for braiding operators,\citep{PhysRevLett.86.268,PhysRevA.73.042313,0268-1242-27-12-124003}
namely the Yang-Baxter equations
\begin{gather}
\begin{gathered}B_{12}B_{23}B_{12}=B_{23}B_{12}B_{23},\\
B_{23}B_{34}B_{23}=B_{34}B_{23}B_{34},
\end{gathered}
\label{eq:5.4}
\end{gather}

\noindent and commutation relations

\begin{align}
\begin{split}\left[B_{12},B_{34}\right]= & 0,\\
\left[B_{12},B_{23}\right]= & \hat{R}_{1}\hat{R}_{3},\\
\left[B_{23},B_{34}\right]= & \hat{R}_{2}\hat{R}_{4}.
\end{split}
\label{eq:5.5}
\end{align}

We also have the non-local braid operators

\begin{equation}
\begin{split}B_{13}= & \exp\left(-\frac{\pi}{4}\hat{R}_{1}\hat{R}_{3}\right),\\
B_{14}= & \exp\left(-\frac{\pi}{4}\hat{R}_{1}\hat{R}_{4}\right),\\
B_{24}= & \exp\left(-\frac{\pi}{4}\hat{R}_{2}\hat{R}_{4}\right),
\end{split}
\label{eq:5.6}
\end{equation}

\noindent connected to the local ones in Eq. (\ref{eq:5.3}) through
the operations

\begin{gather}
\begin{gathered}B_{13}=B_{23}B_{12}B_{23}^{\dagger},\\
B_{14}=B_{34}B_{23}B_{12}B_{23}^{\dagger}B_{34}^{\dagger},\\
B_{24}=B_{34}B_{23}B_{34}^{\dagger}.
\end{gathered}
\label{eq:5.7}
\end{gather}

\noindent The relevant operator to obtain entanglement is $B_{23}$
in Eq. (\ref{eq:5.3}), since it cannot be written as the tensor product
of two $2\times2$ matrices, and therefore is an entangling gate.
This also holds for all three operators in Eq. (\ref{eq:5.7}). $B_{12}$
and $B_{34}$, on the other hand, are separable 

\begin{align}
\begin{split}B_{12}= & \mathbbm{1}_{2}\otimes R_{x}\left(\pi/2\right),\\
B_{34}= & R_{y}\left(\pi/2\right)\otimes\mathbbm{1}_{2},
\end{split}
\label{eq:5.8}
\end{align}

\noindent where $R_{x}\left(\pi/2\right)$ and $R_{y}\left(\pi/2\right)$
are the one-qubit gates (rotation matrices)

\begin{equation}
\begin{split}R_{x}\left(\pi/2\right)= & \exp\left(i\frac{\pi}{4}\sigma^{1}\right),\\
R_{y}\left(\pi/2\right)= & \exp\left(i\frac{\pi}{4}\sigma^{2}\right).
\end{split}
\label{eq:5.9}
\end{equation}

\noindent Thus, leaving out the identity, the braid gates of the model
form the set

\begin{equation}
\left\{ R_{x}\left(\pi/2\right),R_{y}\left(\pi/2\right),B_{23}\right\} .\label{eq:5.10}
\end{equation}

\noindent Acting on the Majorana operators in Eq. (\ref{eq:4.13}),
the braid operators in Eqs. (\ref{eq:5.3}) and (\ref{eq:5.6}) yield

\begin{equation}
B_{pq}\hat{R}_{k}B_{pq}^{\dagger}=\begin{cases}
\hat{R}_{k} & \text{if}\,\,\,k\notin\left\{ p,q\right\} ,\\
\hat{R}_{q} & \text{if}\,\,\,k=p,\\
-\hat{R}_{p} & \text{if}\,\,\,k=q.
\end{cases}\label{eq:5.10-1}
\end{equation}

We also specify the observables $F_{pq}$

\begin{equation}
\begin{array}{ccc}
F_{pq}=-i\hat{R}_{p}\hat{R}_{q}, &  & p<q\end{array},\label{eq:5.12}
\end{equation}

\noindent which are the fermion parity operators for the pair of Majoranas
$pq$, and the total parity operator $Q$ (topological charge)

\begin{equation}
Q=F_{12}F_{34}=-\hat{R}_{1}\hat{R}_{2}\hat{R}_{3}\hat{R}_{4}.\label{eq:5.11}
\end{equation}

\noindent It can be verified that $Q$ commutes with all braid operators
and observables, in compliance with the superselection rules for total
topological charge conservation.\citep{PhysRevA.73.042313}

To complete the model a computational basis needs to be specified.
We choose to fuse the anyons in the pairs 1, 2 and 3, 4, so we consider
the fermionic operators

\begin{align}
\begin{split}f_{12}= & \frac{1}{2}\left(\hat{R}_{1}+i\hat{R}_{2}\right),\\
f_{34}= & \frac{1}{2}\left(\hat{R}_{3}+i\hat{R}_{4}\right),
\end{split}
\label{eq:5.13}
\end{align}

\noindent producing the states

\begin{equation}
\begin{array}{ll}
\left|\bar{0}\bar{0}\right\rangle , & \left|\bar{1}\bar{0}\right\rangle =f_{12}^{\dagger}\left|\bar{0}\bar{0}\right\rangle ,\\
\\
\left|\bar{0}\bar{1}\right\rangle =f_{34}^{\dagger}\left|\bar{0}\bar{0}\right\rangle , & \left|\bar{1}\bar{1}\right\rangle =f_{34}^{\dagger}f_{12}^{\dagger}\left|\bar{0}\bar{0}\right\rangle ,
\end{array}\label{eq:5.14}
\end{equation}

\noindent where $\left|\bar{0}\bar{0}\right\rangle $ is such that
$f_{12}\left|\bar{0}\bar{0}\right\rangle =f_{34}\left|\bar{0}\bar{0}\right\rangle =0$,
and the over bar is used to distinguish them from the canonical states
in Eq. (\ref{eq:4.3}). Explicitly

\begin{equation}
\begin{array}{cc}
\left|\bar{0}\bar{0}\right\rangle =\frac{1}{2}\begin{pmatrix}1\\
-1\\
-i\\
i
\end{pmatrix}, & \left|\bar{1}\bar{0}\right\rangle =\frac{e^{i\frac{\pi}{4}}}{2}\begin{pmatrix}1\\
1\\
e^{-i\frac{\pi}{2}}\\
e^{-i\frac{\pi}{2}}
\end{pmatrix},\\
\\
\left|\bar{0}\bar{1}\right\rangle =\frac{e^{i\frac{\pi}{4}}}{2}\begin{pmatrix}e^{-i\frac{\pi}{2}}\\
-e^{-i\frac{\pi}{2}}\\
1\\
-1
\end{pmatrix}, & \left|\bar{1}\bar{1}\right\rangle =\frac{1}{2}\begin{pmatrix}-i\\
-i\\
1\\
1
\end{pmatrix}.
\end{array}\label{eq:5.15}
\end{equation}

\noindent These states are separable as is readily checked. The first
digit in the kets corresponds to the occupation number of the fermion
operator $f_{12}$, while the second digit to that of the $f_{34}$
operator. This is verified by acting on the basis with the fermion
parity operators in Eq. (\ref{eq:5.12}), giving 

\begin{equation}
\begin{aligned}\begin{split}F_{12}\left|\bar{0}\bar{0}\right\rangle = & \left|\bar{0}\bar{0}\right\rangle ,\\
F_{12}\left|\bar{1}\bar{0}\right\rangle = & -\left|\bar{1}\bar{0}\right\rangle ,\\
F_{12}\left|\bar{0}\bar{1}\right\rangle = & \left|\bar{0}\bar{1}\right\rangle ,\\
F_{12}\left|\bar{1}\bar{1}\right\rangle = & -\left|\bar{1}\bar{1}\right\rangle ,
\end{split}
\end{aligned}
\label{eq:5.15-1}
\end{equation}

\begin{equation}
\begin{aligned}\begin{split}F_{34}\left|\bar{0}\bar{0}\right\rangle = & \left|\bar{0}\bar{0}\right\rangle ,\\
F_{34}\left|\bar{1}\bar{0}\right\rangle = & \left|\bar{1}\bar{0}\right\rangle ,\\
F_{34}\left|\bar{0}\bar{1}\right\rangle = & -\left|\bar{0}\bar{1}\right\rangle ,\\
F_{34}\left|\bar{1}\bar{1}\right\rangle = & -\left|\bar{1}\bar{1}\right\rangle ,
\end{split}
\end{aligned}
\label{eq:5.15-2}
\end{equation}

\noindent with the plus eigenvalue corresponding to the vacant slot
$\bar{0}$ and the minus sign to the occupied state $\bar{1}$. The
total parity operator gives

\begin{align}
\begin{split}Q\left|\bar{0}\bar{0}\right\rangle = & \left|\bar{0}\bar{0}\right\rangle ,\\
Q\left|\bar{1}\bar{1}\right\rangle = & \left|\bar{1}\bar{1}\right\rangle ,\\
Q\left|\bar{0}\bar{1}\right\rangle = & -\left|\bar{0}\bar{1}\right\rangle ,\\
Q\left|\bar{1}\bar{0}\right\rangle = & -\left|\bar{1}\bar{0}\right\rangle .
\end{split}
\label{eq:5.16}
\end{align}

The model is now complete and the system can be initiated in any pair
of the basis states with the same $Q$ parity, due to total parity
conservation. The last two states in Eq. (\ref{eq:5.16}) correspond
to the fusion rule $\sigma\times\sigma=\psi$, while the first ones
to $\sigma\times\sigma=\boldsymbol{1}_{\text{vac}}$ and $\sigma\times\sigma\times\sigma\times\sigma=\boldsymbol{1}_{\text{vac}}$,
respectively. Whatever the initial states are, braiding anyons two
and and three, with the $B_{23}$ operator, produces the states 

\begin{equation}
\begin{split}B_{23}\left|\bar{0}\bar{0}\right\rangle = & \dfrac{1}{\sqrt{2}}\left(\left|\bar{0}\bar{0}\right\rangle +i\left|\bar{1}\bar{1}\right\rangle \right),\\
B_{23}\left|\bar{0}\bar{1}\right\rangle = & \dfrac{1}{\sqrt{2}}\left(\left|\bar{0}\bar{1}\right\rangle -i\left|\bar{1}\bar{0}\right\rangle \right),\\
B_{23}\left|\bar{1}\bar{0}\right\rangle = & \dfrac{1}{\sqrt{2}}\left(-i\left|\bar{0}\bar{1}\right\rangle +\left|\bar{1}\bar{0}\right\rangle \right),\\
B_{23}\left|\bar{1}\bar{1}\right\rangle = & \dfrac{1}{\sqrt{2}}\left(i\left|\bar{0}\bar{0}\right\rangle +\left|\bar{1}\bar{1}\right\rangle \right),
\end{split}
\label{eq:5.17}
\end{equation}

\noindent which conserve total parity and are maximally entangled.
The former is directly seen from Eq. (\ref{eq:5.16}), while the latter
can be established by their Schmidt decomposition, e. g., for $B_{23}\left|\bar{0}\bar{0}\right\rangle $
we have $B_{23}\left|\bar{0}\bar{0}\right\rangle =\frac{1}{\sqrt{2}}\left(\left|0\right\rangle \otimes\left|0\right\rangle +i\left|1\right\rangle \otimes\left|1\right\rangle \right)$,
with $\left|0\right\rangle $, $\left|1\right\rangle $ given in Eq.
(\ref{eq:4.1}). Similar relations hold for the rest of the states
in Eq. (\ref{eq:5.17}). On the other hand, the braid operators $B_{12}$
and $B_{34}$ produce the same state multiplied by a phase of the
type $\exp\left(\pm i\pi/4\right)$ when acting on the basis in Eq.
(\ref{eq:5.15}), as expected from their Abelian nature expressed
in the first relation of Eq. (\ref{eq:5.5}). The states in Eq. (\ref{eq:5.17})
correspond to the fusion rule $\sigma\times\psi=\sigma$. Finally,
we also verify that these maximal entangled states satisfy the Majorana
condition

\begin{equation}
\begin{split}i\gamma^{2}\left(B_{23}\left|\bar{0}\bar{0}\right\rangle \right)^{*}= & -iB_{23}\left|\bar{0}\bar{0}\right\rangle ,\\
i\gamma^{2}\left(B_{23}\left|\bar{0}\bar{1}\right\rangle \right)^{*}= & -B_{23}\left|\bar{0}\bar{1}\right\rangle ,\\
i\gamma^{2}\left(B_{23}\left|\bar{1}\bar{0}\right\rangle \right)^{*}= & -B_{23}\left|\bar{1}\bar{0}\right\rangle ,\\
i\gamma^{2}\left(B_{23}\left|\bar{1}\bar{1}\right\rangle \right)^{*}= & -iB_{23}\left|\bar{1}\bar{1}\right\rangle .
\end{split}
\label{eq:5.19}
\end{equation}

\section{Concluding remarks}

The methods and results presented regarding bispinor solutions to
the massless Dirac equation are of pedagogical value on their own,
and this value can only be enhanced by the connection to QC, e.g.,
after discussing massless bispinors one can readily introduce logical
two-qubit states and entangling gates, or vice-versa. Calculations
in QC could also benefit from the use of relativistic spinors and
the Clifford algebra of the Dirac gamma matrices. Particularly, the
TQC model presented, where operators and states are readily obtained
departing from the set of entangling gates in Eq. (\ref{eq:4.13}),
provides a suitable playground to test and understand how Majorana
zero modes and topological braiding work, both in the technical and
physical assumptions.

\bibliographystyle{apsrev4-2}

\end{document}